\documentclass[twocolumn,floatfix,prb,eqsecnum,showpacs]{revtex4}
\usepackage{graphicx}
\usepackage{amsmath}
\usepackage{bm}
\begin{document}

\title{Electron-electron correlations in a dynamical impurity system with a Fermi edge singularity}

\author{I. Snyman}
\email{izaksnyman1@gmail.com}
\affiliation{School of Physics, University of the Witwatersrand, PO Box Wits, Johannesburg, South Africa}
\affiliation{National Institute for Theoretical Physics, Private Bag X1, 7602 Matieland, South Africa}
\date{December 2012}
\begin{abstract} 
We study spatial correlations in the ground state of a one-dimensional electron gas coupled to
a dynamic quantum impurity. The system displays a non-trivial many-body effect known as the 
Fermi edge singularity: transitions between discrete internal states of the impurity have a power-law
dependence on the internal energies of the impurity states.  We present compact
formulas for the static current-current correlator and the pair correlation function. These
reveal that spatial correlations induced by the impurity decay slowly (as the third inverse power of
distance) 
and have a power-law energy dependence, characteristic of the Fermi edge singularity.
\end{abstract}

\maketitle

\section{Introduction}
 
The situation where a localized scatterer, with 
discreet internal quantum mechanical degrees of freedom, is coupled to an electron gas, is a common
occurrence both in bulk condensed matter
systems\cite{hew} and in nano-meter scale electronic devices.\cite{gei, gor, cro}
Many non-trivial features of these systems can be explained with the aid of 
simple dynamic quantum impurity models.\cite{mou}
For several such models, non-perturbative results 
have been obtained and these results have played a major role in our understanding of strongly 
correlated electron systems. 

In this context, 
a phenomenon known as the Fermi edge singularity\cite{oth90} has been very influential. 
Its essential ingredients are a Fermi gas that is initially in a stationary state
and a local scattering potential that is abruptly switched on at a
time $t_0$. For times larger than $t_0$, the Fermi gas is no longer
in a stationary state. To see just how far from stationary a typical initial state is,
it is useful to consider the overlap between the ground states of the Hamiltonian
before and after $t_0$. It is found that the overlap is zero, and this is known
as the orthogonality catastrophe.\cite{and67}
Since the initial state is far from stationary, the Fermi sea is severely
shaken up;\cite{ada} the local scattering potential creates a multitude of particle hole excitations.

The original theory\cite{mah67, noz69} was formulated to account for singularities in the photo-emission and
-absorption spectra in some metals. For this problem the natural quantities to investigate
are impurity transition rates. These where found to have a power-law dependence on the 
internal energies of the impurity levels. 
It was subsequently shown that the same physics applies to
transport through a barrier containing an impurity\cite{mat}
level, or through a quantum dot.\cite{aba04, hey}
In recent work, non-equilibrium set-ups in which the Fermi sphere is replaced by a non-equilibrium
distribution has been investigated.\cite{muz03,amb04,aba05,sny07,sny08,gut10,bet11}. 
Complications to the basic theory, for instance what happens if the impurity induces resonant
scattering, was also considered.\cite{mkh11}
Apart from impurity transition rates, current
and noise,\cite{mai} as well as the quench dynamics of the electron gas\cite{bra,bet11b} have 
been investigated.

These studies clearly indicate that the dynamical impurity induces significant
correlations in the electron gas. A system that displays the Fermi edge singularity
can therefore be expected to possess a non-trivially correlated ground state. However,
as far as we know, this has not yet been studied. Perhaps the reason is that some common
realizations of the Fermi edge singularity rely on explicitly time dependent Hamiltonians.
In this work we consider a realization with a time independent Hamiltonian. This allows us
to investigate ground state correlations. Correlations in time, such as current noise,\cite{mai} 
have already been studied. However, spatial correlations, a topic that is receiving significant
attention in other impurity systems,\cite{mit,hol,oh} have not yet been addressed. We therefore
do so in this work. 

We consider a set-up in which a localized impurity interacts with one-dimensional electrons in a single
channel with a linear dispersion relation.\cite{she} 
An example of the type of impurity we have in mind, is
the charge qubit formed when a single electron is trapped in a double quantum dot.\cite{elz03,pet04}
Our model can also be realized with a Josephson charge qubit.\cite{gri}
Our main result is a compact formula for the current-current correlator 
$\left<\delta j(x)\delta j(y)\right>$ (Eq.\,\ref{mr}), that is valid at large separations.
The impurity induces no additional correlations between electrons on the same side of the impurity.
For electrons on opposite sides of the impurity, impurity induced correlations decay as $1/|x-y|^3$. 
Owing to the Fermi-edge singularity,
the correlator has a power-law dependence on the internal energies of the impurity.
The correlations persist beyond the range of the interaction between the electron gas and the impurity.
This is in contrast to the expectation values of single particle observables such as the
density $\left<\rho(x)\right>$, which is unaffected by the impurity for $x$ outside the
range of the interaction.
Because of the linear dispersion relation, the current-current correlator is essentially equal
to the pair correlation function $\left<\rho(x)\rho(y)\right>$. This insight allows 
for a simple interpretation of our results. If an electron is detected at $x$ to the
left of the impurity, the likelihood of finding another at $y$ to the right of the impurity
is larger than in the absence of the impurity by an amount $\sim 1/|x-y|^3$ and 
has a power-law dependence on the internal energies of the impurity, characteristic of the
Fermi edge singularity.

We present the work as follows: We define the model by means of its Hamiltonian (Sec. \ref{sham}.)
Since we are dealing with one dimensional electrons with a linear dispersion relation,
it is convenient to bosonize them. 
In Sec.\,\ref{sbos} we collect the relevant bosonization
results. We are interested in the model because it displays a Fermi edge singularity.
We start Sec.\,\ref{sfes} by reviewing how the conventional theory of the Fermi edge
singularity applies to the transition rates between internal states of the impurity. We
also take the first step towards studying ground state properties by considering the probability
to find the impurity in a given internal state, if the system as a whole is in the ground state.
These occupation probabilities turn out to play a central role in determining spatial
correlations between electrons in the presence of the impurity. We show that the Fermi edge
singularity is also manifested in the occupation probabilities. As already stated, our main
aim is to investigate spatial correlations among the electrons. But before doing so, we need to
consider the expectation values of single particle observables. In Sec.\,\ref{scur}
we therefore investigate the average density $\left<\rho(x)\right>$. We show that the
total number of particles displaced by the impurity obeys a generalization of Friedel's sum rule.
(A generalized Friedel sum rule has previously been shown to hold for another dynamic impurity
system, namely the Anderson impurity model.) Our results for the current-current correlator is presented
in Sec.\,\ref{scor} and for the pair correlation function in Sec.\,\ref{spair}. Sec.\,\ref{scon}
contains a self-contained summary of our main result and conclusions. 

\section{Model}
\label{s2}
Our model describes a localized impurity
with an internal state space spanned by the orthonormal basis $B=\left\{\left|\alpha\right>|\alpha=1,\,\ldots,\,M\right\}$. It interacts with
electrons in a single channel with a linear dispersion relation $E_k=k$.
The model is inspired by
experimental setups in which elecrons in a coherent conductor are coupled to a charge
qubit ($M=2$).\cite{sny07,she} 
Since it does not substantially complicate our analysis, we consider 
arbitrary $M$.

\subsection{Hamiltonian}
\label{sham}
The Hamiltonian for our model is $H=H_0+H_T$, where
\begin{align}
&H_0=\sum_{\alpha} H_\alpha\otimes\left|\alpha\right>\left<\alpha\right|,~~H_T=\sum_{\alpha\beta}\gamma_{\alpha\beta}
\left|\alpha\right>\left<\beta\right|,\\
&H_\alpha=\int_{-L/2}^{L/2} dx\,\psi^\dagger(x)\left[-i\partial_x+v_\alpha(x)\right]\psi(x)+\varepsilon_\alpha.
\label{model}
\end{align}
Here $H_\alpha$ describes the fermions when the impurity is
in state $\left|\alpha\right>$ and $\varepsilon_\alpha$ is
the internal energy of the impurity state $\left|\alpha\right>$. 
The fermion creation and annihilation operators $\psi^\dagger(x)$ and 
$\psi(x)$ obey periodic boundary conditions on the interval $(-L/2,L/2]$.  
We will eventually send the system size $L$ to infinity.
The electrostatic
potential $v_\alpha(x)$ depends on the state of
the impurity. Since our aim in this work is to investigate many-body correlations
that persist beyond the range of the interaction between the impurity and the Fermi
gas, it is convenient to assume that all the potentials $v_\alpha$ have bounded support.
We define the scattering region as the shortest interval $l=[x_-,x_+]$ such that 
$v_\alpha(x)=0$ for all $\alpha$ and all $x\not\in l$.
The term $H_T$ induces tunnelling between impurity states. Without loss of generality, we set $\gamma_{\alpha\alpha}=0$. 

The ground state $\left|F_\alpha\right>$ of the fermion Hamiltonian $H_\alpha$ is a Fermi-sea in which all 
(the infinitely many) negative energy orbitals associated with the single particle Hamiltonian $-i\partial_x+v_\alpha(x)$ are
occupied, and all positive energy orbitals are empty. 

The scattering states of  
$-i\partial_x + v_\alpha(x)$ are of the form $\exp[ikx-i\int_{-\infty}^xdx'v_\alpha(x')]$. 
This leads to an identification of 
\begin{equation}
\delta_\alpha=-\frac{1}{2}\int_{-\infty}^\infty dx\, v_\alpha(x)\label{ps}
\end{equation}
as the scattering phase shift associated with the potential $v_\alpha(x)$.
Phase shifts play an important role in impurity physics.

\subsection{Bosonization}
\label{sbos}
The infinite Fermi sea represented by a state such as $\left|F_\alpha\right>$ must be
handled with care. Finite observable quantities are often
represented as the differnece between infinite quantities. In such calculations,
an uncontrolled rearrangement of terms can lead to incorrect answers.\cite{sch59,mat65} 
For one-dimensional Fermi systems, such as the one we are studying, a standard and elegant
method to treat the infinitely deep Fermi sea correctly, is provided by bosonization.\cite{hal81,del98}
In fact, one of the pioneering applications of bosonization was to the Fermi edge singularity.\cite{sch69} 
Through bosonization, our model is mapped onto a spin-boson model\cite{leg87} with a bath that is Ohmic 
at low energies.\cite{she}
For our purposes the following bosonization results are required.
The Fourier components of the density operator 
\begin{equation}
\rho_q=\int_{-L/2}^{L/2}dx\,e^{-iqx}\psi^\dagger(x)\psi(x),
\end{equation}
with $q=2\pi n/L$, $n=0,\,\pm1,\,\pm2,\ldots$ obey bosonic commutation relations
\begin{equation}
[\rho_q,\rho_{q'}]=\frac{Lq}{2\pi}\delta_{q,-q'}.\label{eqa}
\end{equation}
The fermion Hamiltonians $H_\alpha$ can be expressed in terms of $\rho_q$ as $H_\alpha=h_\alpha+E_{0\alpha}$ where
\begin{equation}
h_\alpha=\sum_{q>0} \frac{2\pi}{L}\left(\rho_{-q}+\frac{v_{\alpha,-q}}{2\pi}\right)\left(\rho_q+\frac{v_{\alpha,q}}{2\pi}\right).
\end{equation}
Here $v_{\alpha,q}=\int_{-L/2}^{L/2} dx\,e^{-iqx}v_\alpha(x)$ are the Fourier components of the 
potentials $v_\alpha(x)$ and
\begin{equation} 
E_{0\alpha}=\varepsilon_\alpha+\frac{\rho_0 v_{\alpha,0}}{L}-\sum_{q>0}\frac{2\pi}{L}
\left|\frac{v_{\alpha,q}}{2\pi}\right|^2,\label{egsa}
\end{equation} 
is the ground state energy of $H_\alpha$ as a function of
$\rho_0$, the number of electrons in the system, up to an $\alpha$-independent constant. 
We will work with a fixed number of
particles. The number operator $\rho_0$, and therfore also $E_{0\alpha}$, 
can then be treated as real numbers in stead of operators.
Note that $E_{0\alpha}$ can be varied independently of $h_\alpha$ by adjusting the internal energy 
$\varepsilon_\alpha$ of the impurity state $\alpha$. We define the energy differences 
\begin{equation}
\omega_{\alpha\beta}=E_{0\alpha}-E_{0\beta}.
\end{equation}

For $q>0$, 
\begin{equation}
\rho_q\left|F_\alpha\right>=-\frac{v_{\alpha,q}}{2\pi}\left|F_\alpha\right>,\label{eqb}
\end{equation}
i.e. the Fermi sea $\left|F_\alpha\right>$ is an eigenstate (coherent state) of the bosonic
annihilation operator $\rho_q$.

\subsection{Fermi edge singularity}
\label{sfes}
The Fermi edge singularity manifests as power law singularities in inelastic transition
rates between internal states of the impurity, 
as a function the energy differences $\omega_{\alpha\beta}$. 
Denote the many-body eigenstates of $H_\beta$ by $\left|m_\beta\right>$, 
and let $E_{m\beta}$ be the associated energies. 
Assuming an initial state
$\left|F_\alpha\right>\otimes\left|\alpha\right>$, the total transition rate 
$W_{\beta\alpha}(\omega_{\beta\alpha})$ to configurations with the impurity in state $\left|\beta\right>$,
calculated to second order in $H_T$ using Fermi's golden
rule, is given by
\begin{align}
&W_{\beta\alpha}(\omega_{\beta\alpha})\nonumber\\
&=2\pi|\gamma_{\beta\alpha}|^2\sum_m
\delta(E_{m\beta}-E_{0\alpha})|\left<F_\alpha\right|\left.m_\beta\right>|^2\nonumber\\
&=|\gamma_{\beta\alpha}|^2\sum_m\int_{-\infty}^\infty dt\,e^{i(E_{m\beta}-E_{0\alpha})t}|
\left<F_\alpha\right|\left.m_\beta\right>|^2\nonumber\\
&=|\gamma_{\beta\alpha}|^2\int_{-\infty}^\infty dt\,e^{i\omega_{\beta\alpha} t}
\left<F_\alpha\right|e^{ih_\beta t}
e^{-ih_\alpha t}\left|F_\alpha\right>.\label{eqW}
\end{align}
The canonical theory of the Fermi edge singularity provides an expression
(the so-called closed loop factor)\cite{noz69,aba04}
for the expectation value
$\left<F_\alpha\right|e^{ih_\beta t}e^{-ih_\alpha t}\left|F_\alpha\right>$.
The closed loop factor is analytic in the upper half of the complex plain. For large $|t|$, its
asymptotic behaviour is of the form $\left(i\Lambda t\right)^{-\Delta_{\alpha\beta}^2}$ where
\begin{equation}
\Delta_{\alpha\beta}=-(\delta_\alpha-\delta_\beta)/\pi.\label{rate2}
\end{equation}
A branch cut in the lower half of the complex $t$ plane is understood, and the branch with $\arg(t)=0$
for real positive $t$ is implied. $\Lambda$ is a large energy scale. In our model, with an infinitely deep 
Fermi-sea, it is of the order of the inverse of the range of the interaction of the impurity and the Fermi gas. 
The interpretation is that an interaction with range $\Lambda^{-1}$ can excite particle-hole pairs with energies 
$\sim\Lambda$.
In models where the inverse of the interaction range is larger than the Fermi energy, measured from the
bottom of the Fermi sea, $\Lambda$ is of the order of the Fermi energy. From the asymptotics and analiticity
structure of the closed loop factor, the asymptotic behaviour of the rates $W_{\beta\alpha}$ can be extracted
at small energy differences $\omega_{\beta\alpha}$. 
\begin{equation}
W_{\alpha\beta}(\omega_{\beta\alpha})=\left.|\gamma_{\alpha\beta}|^2 C_{\alpha\beta}\Delta_{\alpha\beta}^2
\theta(\omega)
\left(\frac{\omega}{\Lambda}\right)^{\Delta_{\alpha\beta}^2}
\frac{1}{\omega}\right|_{\omega=-\omega_{\beta\alpha}}.\label{rate1}
\end{equation}
$C_{\alpha\beta}$ is a constant that tends to a value of order one when $\Delta_{\alpha\beta}\to 0$. 
The result for the rate
breaks down at energies $\omega_{\alpha\beta}$ corresponding to excitations with
wave-lengths $\omega_{\alpha\beta}^{-1}$ short enough to resolve the spatial structure of the 
interaction potentials $v_\alpha(x)$ and $v_\beta(x)$. For energies larger than $\Lambda$, the transition
rate becomes exponentially suppressed. There is also another constraint. For perturbation
theory in $\gamma_{\alpha\beta}$ to be valid, the ratio 
$W_{\beta\alpha}(\omega_{\beta\alpha})/\omega_{\alpha\beta}$ must be much smaller than one.\cite{naz09} For 
$\Delta_{\alpha\beta}^2<2$ this imposes the constraint\cite{fn} that 
\begin{equation}
\frac{\omega_{\alpha\beta}}{\Lambda}\gg\left(\frac{C_{\alpha\beta}\Delta_{\alpha\beta}^2|\gamma_{\alpha\beta}|^2}
{\Lambda^2}\right)^{\frac{1}{2-\Delta_{\alpha\beta}^2}}.
\end{equation} 
The point $\omega_{\beta\alpha}=0$ is known as the Fermi edge threshold. More information regarding 
the above
statements can be found in Ref.\,\onlinecite{she}, where we analysed transition rates for essentially
the same model in detail. 
The fact that, for $\Delta_{\beta\alpha}<1$, the transition rate $W_{\omega_{\beta\alpha}}$ becomes
large close to the threshold, while for $\Delta_{\beta\alpha}>1$ it becomes small close to
the threshold, is an instance of what is known as a Schmidt transition.\cite{naz09}

The transition rate discussed above is a dynamical property of the model. However, the Fermi edge
singularity also manifests in ground state properties. Denote the
ground state of the full Hamiltonian by $\left|{\rm GS}\right>$. 
We define $n_\alpha$ as the probability to find the impurity in state $\left|\alpha\right>$ when the system
is in $\left|{\rm GS}\right>$
\begin{equation}
n_\alpha=\left<{\rm GS}\right|\left(\left|\alpha\right>\left<\alpha\right|\right)\left|{\rm GS}\right>.
\end{equation}
Let $\lambda$ 
be the index such that $\left|F_\lambda\right>\otimes\left|\lambda\right>$ is
the ground state of $H_0$, the Hamiltonian in the absence of impurity tunnelling.
It is straight-forward to show that, to second order in $H_T$, $n_\alpha$ with $\alpha\not=\lambda$, is
given by
\begin{equation}
n_\alpha=\int_{-\infty}^0\frac{d\omega}{2\pi}\frac{W_{\alpha\lambda}(\omega)}{(\omega_{\alpha\lambda}-\omega)^2}.
\label{nw}
\end{equation}
(See Appendix \ref{appa} for a derivation.) 
Note that, since $\lambda$ refers to the ground state of $H_0$, 
$\omega_{\alpha\lambda}$ is positive and the denominator of the integrand is therefore always non-zero.
Using the expression of Eq.\,\ref{rate2} for $W_{\alpha\lambda}$ one finds
\begin{equation}
n_\alpha(\omega_{\alpha\lambda})=|\gamma_{\alpha\lambda}|^2 D_{\alpha\lambda}^2
\left(\frac{\omega_{\alpha\lambda}}{\Lambda}\right)^{\Delta_{\alpha\lambda}^2}
\frac{1}{\omega_{\alpha\lambda}^2}.\label{prob}
\end{equation}
Here $D_{\alpha\lambda}$ is a constant of order one (at least for $\Delta_{\alpha\lambda}^2<2$). 
The regime of validity of the above equation
is the same as that of Eq.\,\ref{rate2} for $W_{\alpha\lambda}$.
In particular, for $\Delta_{\alpha\lambda}^2<2$, it becomes invalid at sufficiently small
$\omega_{\alpha\lambda}$ due to a break-down in perturbation theory in $H_T$. For $\Delta_{\alpha\lambda}^2>2$
there are additional contributions that are analytic at small $\omega_{\alpha\lambda}$. These
contributions are however not important for the quantities that we consider in this work, as
they don't give rise to any singular dependence on $\omega_{\alpha\lambda}$. For large 
$\omega_{\alpha\lambda}$ there is a cross-over from the power law of Eq.\,\ref{prob} to 
an inverse square law $n_\alpha\sim(\omega_{\alpha\lambda})^{-2}$.

\section{Current and Friedel sum rule}
\label{scur}
Using the commutation relations between Fourier components of the density operator, the continuity
equation $i[\rho(x),H]=\partial_x j(x)$ is straight-forwardly derived where the current operator is
\begin{equation}
j(x)=\rho(x)+\sum_\alpha\frac{v_\alpha(x)}{2\pi}\left|\alpha\right>\left<\alpha\right|.\label{eqj}
\end{equation}

As a preliminary to our goal of studying correlation functions,
we consider the average current when the system is in an arbitrary stationary state. 
Let $D$ be any density matrix that describes a stationary state of the system, i.e. $[H,D]=0$.
The expectation value of the commutator of any operator $A$ with the Hamiltonian is zero, i.e.
\begin{equation}
\left<[A,H]\right>\equiv{\rm tr}\{D[A,H]\}=0.
\end{equation}
From the continuity equation it then follows that $\partial_x\left<j(x)\right>=0$.
Integrating from $-\infty$ to $x$ and using the explicit expression of Eq.\,\ref{eqj} for $j(x)$, we obtain
\begin{equation}
\left<\rho(x)\right>=\bar{\rho}-\sum_\alpha\frac{v_\alpha(x)}{2\pi}n_\alpha,\label{eqrho}
\end{equation}
where $n_\alpha=\left< \left|\alpha\right> \left<\alpha\right| \right>$ is the probability to find the
impurity in the state $\left|\alpha\right>$ and $\bar{\rho}$ is the density at infinity, which is equal to
the homogeneous density in the absence of the impurity. The total charge $\Delta N$ displaced by the impurity,
is obtained by subtracting $\bar{\rho}$  and integrating over $x$. In terms of the scattering phase shifts
defined in Eq.\,\ref{ps}, we obtain
\begin{equation}
\Delta N=\frac{1}{\pi}\sum_{\alpha=1}^M n_\alpha\,\delta_\alpha.\label{fsr}
\end{equation}
For the case of a static impurity, i.e. $M=1$ and $n_1=1$, the result reduces to Friedel's well-known
sum rule $\Delta N=\delta/\pi$.\cite{hew} Eq.\,\ref{fsr} represents a generalization of Friedel's sum
rule to our dynamic impurity model. That such a generalization exists is not unexpected. 
A classic result in many-body theory is the proof by Langreth\cite{lan} that $\Delta N=\delta/\pi$ holds 
for another dynamic impurity system, namely the Anderson impurity model. 

Before considering correlation functions, we define the current-fluctuation operator $\delta j(x)$
through its Fourier transform $\delta j_q=\int_{-L/2}^{L/2}dx\,e^{-iqx} \delta j(x)$,
where
\begin{align}
\delta j_q&=\sum_\alpha j_{\alpha q}\left|\alpha\right>\left<\alpha\right|-\delta_{q0}j_{\lambda 0},\nonumber\\
j_{\alpha q}&=\rho_q + \frac{v_{\alpha q}}{2\pi}.
\end{align}
This definition is such that the expectation value of $\delta j_q$ with respect to the
non-interacting ground state $\left|F_\lambda\right>\otimes\left|\lambda\right>$ is zero. From
the continuity equation follows that the expectation value of $\delta j_q$ with respect to the 
interacting ground state $\left|{\rm GS}\right>$ is also zero, unless $q=0$. Using the definition of 
$\Delta_{\alpha\lambda}$ in Eq.\,\ref{rate2} and the fact that $\sum_\alpha n_\alpha=1$, one furthermore 
finds
\begin{equation}
\left<{\rm GS}\right|\delta j_q\left|{\rm GS}\right>=\delta_{0q}\sum_\alpha \Delta_{\alpha\lambda}n_\alpha.
\end{equation}
For a finite interaction strength between the impurity and the Fermi gas, the relative phase shifts
$\Delta_{\alpha\lambda}$ are finite, and hence in the thermodynamic limit $L\to\infty$, the Fourier
transformed fluctuations operator $\delta j(x)$ has a zero expectation with respect to the
full ground state
\begin{equation}
\left<{\rm GS}\right|\delta j(x)\left|{\rm GS}\right>=\frac{1}{L}\sum_\alpha \Delta_{\alpha\lambda}n_\alpha
\to \left. 0\right|_{L\to\infty}.
\end{equation}

\section{Current-Current correlators}
\label{scor}
The similarity of the generalized Friedel sum rule of Eq\,\ref{fsr} to the static impurity result,
$\Delta N=\delta/\pi$,
begs the following question: Given the probabilities $n_\alpha$, can the effect of the impurity on
the electron gas perhaps be accounted for by a static
effective potential $v_{\rm eff}=\sum_{\alpha}n_\alpha\,v_\alpha(x)$ in stead of a dynamic impurity? 
In this section, we show that the answer is ``No''. 
We show this by examening the ground state current-current correlator
$\left<{\rm GS}\right|\delta j(x)\delta j(y)\left|{\rm GS}\right>$, which 
can in principle be extracted from the statistics of
repeated experiments in which the current is measured. (See Appendix \ref{appb}.)


\subsection{Zero'th order}
\label{szero}
We write
\begin{equation}
\left<{\rm GS}\right|\delta j(x)\delta j(y)\left|{\rm GS}\right>=c_0(x,y)+c_2(x,y)+
\mathcal O(\gamma^4),\label{eq15}
\end{equation}
where $c_n$ is of order of $n$ in $\gamma$.
The zero-order term 
\begin{equation}
c_0(x,y)=\left<F_\lambda\right|\delta j(x)\delta j(y)\left|F_\lambda\right>,\label{eq16}
\end{equation}
becomes singular when $x\to y$. To deal with this singularity, it is convenient to
define a regularised current operator
\begin{eqnarray}
\delta j_r(x)&=&\frac{1}{L}\sum_q e^{iqx} e^{-r |q|}\delta j_q\nonumber\\
&\simeq&\int_{-L/2}^{L/2} \frac{dx'}{\pi} \frac{r}{(x-x')^2+r^2} \delta j(x),\label{eq17}
\end{eqnarray}
the last line being valid for $|x|/L\ll 1$. At the end of the calculation the limit $r\to 0^+$ is taken and
results independent of the regularization is obtained as long as $x\not= y$.
From Eqs.\,\ref{eqa} and \ref{eqb} follows that
\begin{equation}
\left<F_\lambda\right|\delta j_p \delta 
j_q\left|F_\lambda\right>=\theta(p)\delta_{p,-q}\frac{Lp}{2\pi}.\label{eq18}
\end{equation}
Performing the inverse Fourier transform we then find
\begin{align}
&\left<F_\lambda\right|\delta j(x)\delta j(y)\left|F_\lambda\right>\nonumber\\
&=\frac{-i}{2\pi L}\partial_x\sum_{p>0} e^{ip(x-y+2ir)}\nonumber\\
&=\frac{-i}{2\pi L}\partial_x\left[1-e^{i\frac{2\pi}{L}(x-y+2ir)}\right]^{-1}.\label{eq19}
\end{align}
Taking the limits $L\to\infty$ and $r\to0^+$ we find
\begin{equation}
c_0(x,y)=-\frac{1}{(2\pi)^2}\frac{1}{(y-x)^2}.\label{eq20}
\end{equation}
This result is independent of $v_\lambda(x)$ and represents the exact result for
a static impurity. 

\subsection{Secdond order}
\label{stwo}
We obtain the second order correction $c_2(x,y)$ to the current-current correlator 
by expanding $\left|{\rm GS}\right>$ in Eq.\,\ref{eq15} to
second order in $\gamma$. The perturbation expansion of eigenstates
can be obtained from a perturbation expansion of the interaction picture time evolution operator describing the
situation in which the perturbation is switched on adiabatically [Gell-mann Low theorem]. 
Thus, to second order in the $\gamma$'s
\begin{widetext}
\begin{eqnarray}
\left|{\rm GS}\right>&=&\left|F_\lambda\right>\otimes\left|\lambda\right>
+\sum_{\alpha}\left|\psi_\alpha\right>\otimes\left|\alpha\right>+\sum_{\alpha_1\alpha_2}\left|\psi_{\alpha_1\alpha_2}\right>\otimes\left|\alpha_2\right>\label{gsexp}\\
\left|\psi_\alpha\right>&=&-i\sum_{\alpha}\gamma_{\alpha\lambda}\int_{-\infty}^0dt\,e^{\eta t}Q_{\alpha\lambda}(t)\left|F_\lambda\right>,\label{xe1}\\
\left|\psi_{\alpha_1\alpha_2}\right>&=&-\sum_{\alpha_1\alpha_2}\gamma_{\alpha_2\alpha_1}\gamma_{\alpha_1\lambda}\int_{-\infty}^0dt_2\int_{-\infty}^{t_2}
dt_1\,e^{\eta(t_1+t_2)} Q_{\alpha_2\alpha_1}(t_2)Q_{\alpha_1\lambda}(t_1)\left|F_\lambda\right>\otimes\left|\alpha_2\right>.
\label{ex2}
\end{eqnarray}  
\end{widetext}
Here $\eta$ is a small positive constant and the limit $\eta\to0^+$ must be taken after the expectation value 
in Eq.\,\ref{eq15} is evaluated. The operators $Q_{\alpha\beta}(t)$ are defined as
\begin{equation}
Q_{\alpha\beta}(t)=e^{iH_\alpha t}e^{-iH_\beta t}.
\end{equation}
We also define the expectation value
\begin{eqnarray}
P_{\alpha\beta}(t)& \equiv &\left<F_\lambda\right|Q_{\alpha\beta}(t)\left|F_\lambda\right>\nonumber\\
&=&e^{i(\omega_{\alpha\lambda}-\omega_{\beta\lambda})t}\left<F_\lambda\right|e^{ih_\alpha t}e^{-ih_\beta t}\left|F_\lambda\right>.
\label{eqp}
\end{eqnarray}
It is convenient to perform a Fourier transform
\begin{equation}
c_2(p,q)=\int_{-L/2}^{L/2}dx\int_{-L/2}^{L/2}dy\,e^{-ipx}e^{-iqy}c_2(x,y).\label{eq21}
\end{equation}
Expanding the ground state $\left|{\rm GS}\right>$ in $\gamma$ as in Eq.\,\ref{gsexp}, we obtain
\begin{align}
c_2(p,q)=&\sum_{\alpha}c_2^{\alpha}(p,q),\nonumber\\
c_2^{\alpha}(p,q)=&\left<\psi_\alpha\right|\delta j_{\alpha p}\delta j_{\alpha q}
\left|\psi_\alpha\right>+\left<F_\lambda\right|
\delta j_{\lambda p}\delta j_{\lambda q}\left|\psi_{\alpha\lambda}\right>\nonumber\\
&+\left<\psi_{\alpha\lambda}\right|
\delta j_{\lambda p}\delta j_{\lambda q}\left|F_\lambda\right>.\label{calp}
\end{align}

We now substitute $\left|\psi_{\alpha}\right>$ and $\left|\psi_{\alpha\beta}\right>$ from Eqs.\,\ref{xe1} and \ref{ex2} into Eq.\,\ref{calp}.
The individual terms in Eq.\,\ref{calp} are evaluated by exploiting the bosonic commuators $[\delta j_{\alpha q},\delta j_{\beta q'}]=
Lq\delta_{q,-q'}/2\pi$ obeyed by the $j_{\alpha q}$
operators, (cf. Eq.\,\ref{eqa}). For $q>0$, the $\delta j_{\alpha q}$ correspond to bosonic annihilation operators, with corresponding creation operators $\delta j_{\alpha q}^\dagger=\delta j_{\alpha-q}$. 
The state $\left|F_\lambda\right>$
is a coherent state, i.e. an eigenstate of the annihilation operators, (cf. Eq.\,\ref{eqb})
\begin{equation}
\delta j_{\alpha q}\left|F_\lambda\right>=\Delta_{\alpha\lambda}^q\left|F_\lambda\right>,
\hspace{1cm}\Delta_{\alpha\lambda}^q=\frac{v_{\alpha q}-v_{\lambda q}}{2\pi}.
\end{equation}
The commutator of $\delta j_{\alpha q}$ with $Q_{\beta\gamma}(t)$ 
is easily calculated using  the BCH identity. The result is
\begin{equation}
[\delta j_{\alpha q},Q_{\beta\gamma}]=(e^{iqt}-1)\Delta_{\beta\gamma}^q Q_{\beta\gamma}(t).
\end{equation}
After some algebra, we obtain the expression $c_2^\alpha(p,q)=C_2^{\alpha}(p,q)+C_2^{\alpha}(-p,-q)^*$ where
\begin{widetext}
\begin{align}
C_2^\alpha(p,q)=\lim_{\eta\to0^+}|\gamma_{\alpha\lambda}|^2\Delta_{\alpha\lambda}^p\Delta_{\alpha\lambda}^q\int_{-\infty}^0dt_1
\int_{-\infty}^{t_1}dt_2\,&\left\{e^{i(p+q)t_1}+\theta(p)e^{iqt_1}\left[e^{ipt_2}-e^{ipt_1}\right]+
\theta(q)e^{ipt_1}\left[e^{iqt_2}-e^{iqt_1}\right]\right\}\nonumber\\
&\times e^{\eta(t_1+t_2)}P_{\alpha\lambda}(t_2-t_1)
\end{align}
We change integration variables to $t=t_1$ and $\tau=t_2-t_1$ and perform the $t$ integral to obtain
\begin{equation}
C_2^\alpha(p,q)=\lim_{\eta\to0^+}\frac{|\gamma_{\alpha\lambda}|^2}{2\eta+i(p+q)}\Delta_{\alpha\lambda}^p\Delta_{\alpha\lambda}^q\int_{-\infty}^0d\tau
\left[1+\theta(p)\left(e^{ip\tau}-1\right)+
\theta(q)\left(e^{iq\tau}-1\right)\right]e^{\eta\tau}P_{\alpha\lambda}(\tau).
\end{equation}
From Eq.\,\ref{eqp} follows that $P_{\alpha\lambda}(t)$ 
is related to the transition rate $W_{\alpha\lambda}$ by a Fourier transform.
\begin{equation}
|\gamma_\alpha\lambda|^2P_{\alpha\lambda}(t)=\int_{-\infty}^0\frac{d\omega}{2\pi}
e^{i(\omega_{\alpha\lambda}-\omega)t}W_{\alpha\lambda}(\omega).
\label{eqft}
\end{equation} 
Substituting for $|\gamma_{\alpha\lambda}|^2P_{\alpha\lambda}(\tau)$ from Eq.\,\ref{eqft}, performing the $\tau$ integral and 
taking the $\eta\to0^+$ limit, we find
\begin{equation}
C_2^\alpha(p,q)=\frac{\Delta_{\alpha\lambda}^p\Delta_{\alpha\lambda}^q}{p+q}\int_{-\infty}^0\frac{d\omega}{2\pi}W_{\alpha\lambda}(\omega)
\left\{\frac{1}{\omega-\omega_{\alpha\lambda}}+\theta(p)\left[\frac{1}{\omega-\omega_{\alpha\lambda}-p}-\frac{1}{\omega-\omega_{\alpha\lambda}}\right]
+\theta(q)\left[\frac{1}{\omega-\omega_{\alpha\lambda}-q}-\frac{1}{\omega-\omega_{\alpha\lambda}}\right]\right\},
\end{equation}
and hence 
\begin{equation}
c_2^\alpha(p,q)=\frac{\Delta_{\alpha\lambda}^p\Delta_{\alpha\lambda}^q}{p+q}\int_{-\infty}^0\frac{d\omega}{2\pi}
W_{\alpha\lambda}(\omega)
\left\{{\rm sign}(p)\left[\frac{1}{\omega-\omega_{\alpha\lambda}-|p|}-\frac{1}{\omega-\omega_{\alpha\lambda}}\right]
+{\rm sign}(q)\left[\frac{1}{\omega-\omega_{\alpha\lambda}-|q|}-\frac{1}{\omega-\omega_{\alpha\lambda}}\right]
\right\}.
\end{equation}
Fourier transforming back to $c_2^\alpha(x,y)$ we obtain
\begin{equation}
c_2^\alpha(x,y)=\int_{-\infty}^0 \frac{d\omega}{2\pi}W_{\alpha\lambda}(\omega)\int_{-\infty}^0dz\int_0^\infty dz'
\left[\Delta_{\alpha\lambda}(x+z)\Delta_{\alpha\lambda}(y+z')+\Delta_{\alpha\lambda}(x+z')\Delta_{\alpha\lambda}(y+z)\right]h_\alpha(z'-z),
\label{cxy1}
\end{equation}
\end{widetext}
where 
\begin{equation}
\Delta_{\alpha\lambda}(x)=\int_{-\infty}^\infty 
\frac{dq}{2\pi}\,e^{iqx}\Delta_{\alpha\lambda}^q=\frac{v_\alpha(x)-v_\lambda(x)}{2\pi},
\end{equation}
and
\begin{equation}
h_\alpha(z)=\int_0^\infty \frac{dq}{\pi}\,\sin(qz)\left(\frac{1}{\omega_{\alpha\lambda}-\omega}-\frac{1}{\omega_{\alpha\lambda}
-\omega+|q|}\right).
\end{equation}
For large $z$\cite{abr} 
\begin{equation}
h_\alpha(z)\simeq\frac{2}{\pi}\frac{1}{(\omega_{\alpha\lambda}-\omega)^3z^3}=\frac{-1}{\pi}
\partial_{\omega_{\alpha\lambda}}\frac{1}{(\omega_{\alpha\lambda}-\omega)^2z^3}.
\label{lzh}
\end{equation}

Since $\Delta_{\alpha\lambda}(x)$ is zero for $x\not\in l$, i.e. outside the scattering region, the first term in the square brackets
in Eq.\,\ref{cxy1} gives a non-zero contribution only when $x>x_-$ and $y<x_+$. The second term in the square brackets on the 
other hand only gives a non-zero contribution when $x<x_+$ and $y>x_-$. Thus for $x$ and $y$ both to the left ($<x_-$) or both
to the right ($>x_+$) of the scattering region, $c_2^\alpha(x,y)$ is zero. It can be shown that 
this statement is true to all orders in $H_T$.

For $x$ and $y$ on opposite sides of the scattering region, and such that $|x-y|\gg x_+-x_-$, we can use the large $z$ expansion of
$h_\alpha(z)$ (cf. Eq.\,\ref{lzh}). For $|z|\gg x_+-x_-$, $h_\alpha(z)$ is a slowly varying function on the scale of $x_+-x_-$. 
The leading order behavior in $|x-y|$ of 
$c_2^\alpha(x,y)$ can be obtained by evaluating $h_\alpha(z'-z)$ at $|x-y|$
\begin{equation}
c_2^\alpha(x,y)\simeq-\frac{1}{\pi}(\Delta_{\alpha\lambda})^2\frac{1}{|x-y|^3}\partial_{\omega_{\alpha\lambda}}
\int_{-\infty}^0\frac{d\omega}{2\pi}
\frac{W_{\alpha\lambda}(\omega)}{(\omega_{\alpha\lambda}-\omega)^2},
\end{equation}
where $\Delta_{\alpha\lambda}\equiv\int_{-l}^l dx\, \Delta_{\alpha\lambda}(x)$. Substituting into the above 
the result of Eq.\,\ref{nw} where the rate $W_{\alpha\lambda}$ was related to the occupation 
probability $n_\alpha$ of impurity level $\alpha$ we obtain the simple final result
\begin{equation}
c_2^\alpha(x,y)\simeq-\frac{1}{\pi}(\Delta_{\alpha\lambda})^2\frac{1}{|x-y|^3}
\partial_{\omega_{\alpha\lambda}}n_\alpha(\omega_{\alpha\lambda}),\label{mr}
\end{equation}
for $\min\{x,y\}\ll x_-$ and $\max\{x,y\}\gg x_+$.

For $\omega_{\alpha\lambda}$ not too large, i.e. not too far from the Fermi edge threshold,
the energy dependence of the correlator can be obtained from the expression (Eq.\,\ref{prob})
for $n_\alpha$ that we discussed in the section on the Fermi edge singularity. This leads
to
\begin{equation}
c^\alpha_2(x,y)\sim\left(\omega_{\alpha\lambda}\right)^{\Delta_{\alpha\lambda}^2-3}.
\end{equation}
For $\Delta_{\alpha\lambda}^2<3$ the result diverges close to the Fermi edge threshold.
As with the rate $W_{\alpha\lambda}$ and the occupation probability $n_\alpha$,
the divergence signals a breakdown of our expansion in $\gamma$.
Understanding this regime is an open problem that we are currently working on.

\section{Pair-correlation function}
\label{spair}
Although it can be measured directly (see Appendix \ref{appb}), the current-current correlator 
that we studied in the previous subsection is a rather abstract quantity. In this subsection
we therefore relate it to another correlator, the pair correlation function, that has an
appealingly straight forward interpretation. The pair correlation function is defined as
\begin{equation}
g(x,y)=\left<{\rm GS}\right|\psi^\dagger(y)\psi^\dagger(x)\psi(x)\psi(y)\left|{\rm GS}\right>.
\end{equation}
It measures the likelihood of finding an electron at $x$ together with another electron at $y$.
For $x\not= y$ it can also be written as 
$g(x,y)=\left<{\rm GS}\right|\rho(x)\rho(y)\left|{\rm GS}\right>$. Since we have $j(x)=\rho(x)$
outside the scattering region (see Eq.\,\ref{eqj}),
our result for the current-current correlator can be used to obtain
a formula for $g(x,y)$ away from the scatterer, namely
\begin{equation}
g(x,y)=\bar{\rho}^2-\frac{1}{(2\pi)^2}\frac{1}{(x-y)^2}+\sum_\alpha c_2^\alpha(x,y)+\mathcal O(\gamma^4).
\label{eqg}
\end{equation}
The negative sign in front of the second term is due to Fermi statistics: Given that there is
an electron at $x$, the likelihood of finding another electron at nearby position $y$ is less than
for uncorrelated distinguishable particles. The term $c_2(x,y)$, as well as the higher 
order corrections, are zero if $x$ and $y$ refer to points on the same side of the
scatterer, i.e. if $x,\,y<x_-$ or $x,\,y>x_+$. 

For $\max\{x;y\}\gg x_+$ and $\min\{x;y\}\ll x_-$, the approximate result of Eq.\,\ref{mr}
may be used for $c_2^\alpha$. The sign of $c_2^\alpha(x,y)$ is determined by 
$\left.-\partial_\omega n_\alpha(\omega)\right|_{\omega=\omega_{\alpha\lambda}}$.
It is intuitively plausible that the probability $n_\alpha$
is a decreasing function of $\omega_{\alpha\lambda}$: the higher the excitation energy, the less likely
it is to find the state $\left|\alpha\right>$ occupied. 
Refering back to Eq. \ref{prob} in our discussion of the Fermi edge singularity, we see that this is
indeed the case not too far from the Fermi edge threshold, provided $\Delta_{\alpha\lambda}^2<2$.
Exact results for specific model interactions indicate that $n_\alpha$ remains a decreasing function
of $\omega_{\alpha\lambda}$ also when $\omega_{\alpha\lambda}$ is further away from the threshold 
or $\Delta_{\alpha\lambda}>2$. This leads to the conclusion that $c_2^{\alpha}(x,y)$ is
always positive.
Thus in the presence of the impurity, the pair correlation function is larger than
in the absence of the impurity. 
This a truly two-particle correlation effect.
It cannot be accounted for by an increase in density around the impurity, for 
the following reasons.
The increase in pair correlations occurs outside the scattering region, 
where $\left<{\rm GS}\right|\rho(x)\left|{\rm GS}\right>$ is unaffected by the presence of the impurity.
Also, the pair correlation
function increases regardless of whether the impurity attracts or repel electrons.
Finally, the increase in the pair correlation function depends on distance $x-y$ 
rather than on the distance from $x$ or $y$ to the impurity.

\section{Conclusions}
\label{scon}
We studied a model in which a dynamical quantum impurity is coupled to a degenerate electron gas.
The probability $n_\alpha$  find the impurity in excited state $\left|\alpha\right>$ if the system as
a whole is in the ground state, has an energy dependence (cf. Eq.\,\ref{prob})
\begin{equation}
n_\alpha\sim(\omega_{\alpha\lambda})^{\Delta_{\alpha\lambda}^2-2},\label{c1}
\end{equation}
where $\omega_{\alpha\lambda}$ is the internal energy of impurity state $\left|\alpha\right>$
minus the threshold energy of the Fermi edge singularity, and $\Delta_{\alpha\lambda}$ is
the relative scattering phase shift that measures the strength of the interaction between impurity
state $\left|\alpha\right>$ and the Fermi gas. The power-law in Eq.\,\ref{c1} is
due to the Fermi edge singularity, a phenomenon that is usually discussed in the
context of impurity transition rates.  We note that the divergence of $n_\alpha$ at the Fermi
edge threshold for $\Delta_{\alpha\lambda}<2$ signals a breakdown in perturbation theory in
the tunnelling matrix elements $\gamma_{\alpha\beta}$ of the impurity.

Eq.\,\ref{c1} establishes that the Fermi edge singularity manifests itself in ground state properties
of the impurity. How does it manifest itself in properties of the electron gas? We found that, for
any stationary state, the average density of the electron gas is (cf. Eq.\,\ref{eqrho})
\begin{equation}
\left<\rho(x)\right>=\bar{\rho}-\sum_\alpha\frac{v_\alpha(x)}{2\pi}n_\alpha,
\end{equation}
where $\bar{\rho}$ is the homogeneous density in the absence of the impurity and $v_\alpha(x)$
is the electrostatic potential that the electron gas is subjected to when the impurity is in state 
$\left|\alpha\right>$. 
This leads to a generalized Friedel sum rule (cf. Eq.\,\ref{fsr}) $\Delta N=\sum_\alpha n_\alpha 
\delta_\alpha/\pi$ for the number of particles displaced by the impurity, where  $\delta_\alpha$
is the scattering phase shift associated with $v_\alpha(x)$. 
Two important points about these results are that (1) the
effect of the Fermi edge singularity is contained in the occupation probabilities $n_\alpha$;
and (2) the average density is unaffected by the impurity in regions where the
potentails $v_\alpha(x)$ are zero, i.e. outside the scattering region.

Having investigated the expectation values of single particle observables, the next logical step
is to look at correlation functions. We therefore calculated the static current-current correlator
$\left<{\rm GS}\right|\delta j(x)\delta j(y)\left|{\rm GS}\right>$ to second order in the impurity
tunneling amplitudes $\gamma_{\alpha\beta}$. The zero-order result, which is also the full answer
in the case of a static impurity) is (cf. Eq.\,\ref{eq20})
\begin{equation}
c_0(x,y)=-\frac{1}{(2\pi)^2(x-y)^2}.\label{c2}
\end{equation}
We obtained an expression for the second order correction $c_2(x,y)$
to this result (Eq.\,\ref{cxy1}) that is valid for all $x$ and $y$. It is zero for $x$ and $y$ both
to the left or both to the right of the scattering region. Taking $x$ and $y$ on different
sides of the scattering region, and the distance $|x-y|$ large, we obtained the compact formula
\begin{equation}
c_2(x,y)\simeq-\frac{1}{\pi|x-y|^3}\sum_\alpha\left(\Delta_{\alpha\lambda}\right)^2
\partial_{\omega_{\alpha\lambda}}n_\alpha(\omega_{\alpha\lambda}).\label{c3}
\end{equation}
Thus, current-current correlations induced by the impurity
show a slow (power-law) decay as a function of distance. These correlations are sub-leading: at
large distances, correlations that are present also in the absence of the impurity decay more slowly
(second vs. third inverse power of distance). However, the impurity-induced correlations can be detected
by varying the impurity parameters, as this leaves the leading order correlations unaffected.
Due to the appearance of $n_\alpha$ in Eq.\,\ref{c3}, the impurity
induced correlations have a power law energy dependence 
$\sim(\omega_{\alpha\lambda})^{\Delta_{\alpha\lambda}^2-3}$. Thus correlations are also sensitive to
the Fermi edge singularity. The divergence at the Fermi edge threshold for $\Delta_{\alpha\lambda}^2<3$
again signals the breakdown of perturbation theory in $\gamma_{\alpha\lambda}$. Understanding correlations
in the regime $(\Delta_{\alpha\lambda})^2<3$ and $\omega_{\alpha\lambda}\to 0$ is an open problem
we are currently working on. In the regime where the Fermi edge singularity result for $n_\alpha$
(Eq.\,\ref{c1}) is valid, the correction $c_2(x,y)$ is positive. We have argued that $n_\alpha$ is
always a decreasing function of $\omega_{\alpha\lambda}$. Thus we always expect a positive correction
$c_2(x,y)$.

Outside the scattering region, the current-current correlator is simply related to the pair 
correlation function $g(x,y)=\left<{\rm GS}\right|\rho(x)\rho(y)\left|{\rm GS}\right>$ as 
\begin{equation}
g(x,y)=\bar{\rho}^2
-\left<{\rm GS}\right|\delta j(x)\delta j(y)\left|{\rm GS}\right>.
\end{equation}
The pair correlation function measures the likelihood to find an electron at $x$ together with 
another electron at $y$. The positivity of $c_2(x,y)$ implies that given an electron at $x$,
the likelihood of finding another one at $y$ in the presence of the impurity increases
by an amount proportional to $1/|x-y|^3$, for $x$ and $y$ on opposite sides of the scattering region,
compared to when the impurity is absent. It is important to note that the increase occurs outside
the scattering region, where the average density of electrons is unaffected by the impurity.

\appendix

\section{The relation between impurity transition rates and ground state occupation probabilities}

\label{appa}
In this appendix we derive a relation (Eq.\,\ref{rate1}) between the ground state occupation
probability $n_\alpha$ and the transition rate $W_{\alpha\lambda}$ of the impurity.
The starting point is the perturbation expansion of Eqs.\,\ref{gsexp} and 
\ref{xe1} for the ground state. From these equations follow that,
to second order in the $\gamma$'s, $n_\alpha$, with $\alpha\not=\lambda$, is given by
\begin{equation}
n_\alpha=\lim_{\eta\to0^+}|\gamma_{\alpha\lambda}|^2\int_{-\infty}^0dt\int_{-\infty}^0dt'\,e^{\eta(t+t')}
P_{\alpha\lambda}(t-t').
\end{equation}
By changing integration variables to $T=(t+t')/2$ and $\tau=t-t'$, performing the $T$-integral we then obtain
\begin{equation}
n_\alpha=\lim_{\eta\to0^+}\frac{|\gamma_{\alpha\lambda}|^2}{2\eta}\int_{-\infty}^\infty d\tau\,
e^{-\eta|\tau|}P_{\alpha\lambda}(\tau).\label{eqnii}
\end{equation}
Substituting for $|\gamma_{\alpha\lambda}|^2P_{\alpha\lambda}(t)$ from (Eq.\,\ref{eqft}) allows us
to perform the $\tau$-integral. Taking the $\eta\to 0^+$ limit, we obtain
\begin{equation}
n_\alpha=\int_{-\infty}^0\frac{d\omega}{2\pi}\frac{W_{\alpha\lambda}(\omega)}{(\omega_{\alpha\lambda}-\omega)^2}.
\end{equation}

\section{Measurability of the current-current correlator}
\label{appb}

We present here what we find to be conceptually the simplest scheme to measure 
the current-current correlator. Other schemes may be more general or more practical.
The Hamiltonian of Eq.\,\ref{model} describes right-moving electrons propagating along the real
line. A straight-forward realization of such a conductor is a quantum Hall edge state.
However, another realization is a semi-infinite wire running from $x=-\infty$ to $x=0$, 
containing both left- and right moving
electrons. In this realization, the fermion creation operators $\psi^\dagger(x)$ and 
$\psi^\dagger(-x)$ both create an electron at a position $-|x|$, while the sign of $x$ determines
whether a left-mover (positive sign) or a right-mover (negative sign) is created.
The physically measurable current at $-|x|$ corresponds to the operator
\begin{equation}
J(-|x|)=j(-|x|)-j(|x|),\label{j}
\end{equation} 
where $j(x)$ is the current operator defined in Eq.\,\ref{eqj}. 
In other words, the physically measured
current equals the current of right-movers minus the current of left-movers.

In spectral representation, we write $J(-|x|)$ as
\begin{equation}
J(-|x|)=\sum_m J_m \left|m\right>\left<m\right|,
\end{equation}
where $J_m$ and $\left|m\right>$ are the many-body eigenvalues and eigenstates of $J(-|x|)$.

By performing current measurements at $-|x|$ on a sufficient number of identical systems
prepared in the ground state, the probability density 
\begin{equation}
P(J)\equiv\sum_m \delta(J_m-J)\left|\left<GS\right|\left.m\right>\right|^2,\label{pj}
\end{equation} 
for an outcome $J$ can approximately be determined. Given $P(J)$, its second moment
\begin{equation}
M_2(-|x|)\equiv\int_{-\infty}^\infty dJ\, J^2 P(J),
\end{equation}
can be extracted. From the definitions of $J(-|x|)$ and $P(J)$ follows that
\begin{align}
M_2(-|x|)=&\left<j(|x|)^2\right>+\left<j(-|x|)^2\right>\nonumber\\
&-\left<j(-|x|)j(|x|)\right>-\left<j(|x|)j(-|x|)\right>.
\end{align}
In the above expression, expectation values are with respect to the ground state.

The expectation value $\left<j(x)^2\right>$ turns out to be position-independent for $x$
outside the range of the impurity interaction, i.e. $x\not\in l$. The proof is as follows:
\begin{eqnarray}
\partial_x\left<j(x)^2\right>&=&\left<\partial_xj(x)j(x)\right>+\left<j(x)\partial_xj(x)\right>\nonumber\\
&=&i\left<[\rho(x),H]j(x)\right>+i\left<j(x)[\rho(x),H]\right>\nonumber\\
&=&i\left<[j(x),H]j(x)\right>+i\left<j(x)[j(x),H]\right>\nonumber\\
&=&0.
\end{eqnarray}
In the second line we exploited the continuity equation, while in the third line
we used the fact that 
$\rho(x)=j(x)$ (cf. Eq. \ref{eqj}) for $x\not\in l$.
The last line is obtained by noting that the expectation value is with respect to $\left|{\rm GS}\right>$
which is an eigenstate of $H$.

Thus, $M_2(-|x|)$ can be measured, and is given by
\begin{equation}
M_2(-|x|)=M+\left<j(-|x|)j(|x|)\right>+\left<j(|x|)j(-|x|)\right>,
\end{equation}
where $M$ is independent of $x$. Our main result (Eq.\,\ref{mr}) then translates into a prediction for the $x$ dependent
part of $M_2$:
\begin{equation}
M_2(-|x|)-M=-\frac{1}{4\pi|x|^3}\sum_{\alpha}(\Delta_{\alpha\lambda})^2
\partial_{\omega_{\alpha\lambda}}n_\alpha(\omega_{\alpha\lambda}).
\end{equation}

\acknowledgments

This research was supported by the National Research Foundation (NRF) of South Africa.


\begin{thebibliography}{99}
\bibitem{fn} The stated condition on $\omega_{\alpha\lambda}$ is discussed in detail in Sec. VII B of
Re.\,\onlinecite{leg87} in the context of the spin-boson model with an Ohmic bath.
\end{thebibliography}
\end{document}